# Network structure exploration via Bayesian nonparametric models

Yi Chen[1], Xiao-long Wang[1,2], Xin Xiang[1], Bu-zhou Tang[1], Qing-cai Chen[1], Bo Yuan[1], Jun-zhao Bu[1]

[1](*Key Laboratory of Network Oriented Intelligent Computation, Shenzhen Graduate School, Harbin Institute of Technology, Shenzhen 518055, China*)

[2](*School of Computer Science and Technology, Harbin Institute of Technology, Harbin 150001, China*)

**Abstract**

Complex networks provide a powerful mathematical representation of complex systems in nature and society. To understand complex networks, it is crucial to explore their internal structures, also called structural regularities. The task of network structure exploration is to determine how many groups in a complex network and how to group the nodes of the network. Most existing structure exploration methods need to specify either a group number or a certain type of structure when they are applied to a network. In the real world, however, not only the group number but also the certain type of structure that a network has are usually unknown in advance. To automatically explore structural regularities in complex networks, without any prior knowledge about the group number or the certain type of structure, we extend a probabilistic mixture model that can handle networks with any type of structure but needs to specify a group number using Bayesian nonparametric theory and propose a novel Bayesian nonparametric model, called the Bayesian nonparametric mixture (BNPM) model. Experiments conducted on a large number of networks with different structures show that the BNPM model is able to automatically explore structural regularities in networks with a stable and state-of-the-art performance.

## Introduction

Complex networks provide a powerful mathematical representation of complex systems throughout many disciplines in nature and society, including informatics[1,2], biology[3,4,5], sociology[6,7,8], ecology[9,10] and engineering[11]. For example, complex networks have been used to represent citations between papers[2], interactions between proteins[5], friendships between people[8], predator-prey interactions between species[10], and physical connections between electronic components[11]. A complex network is a set of nodes (also called vertices) with edges between them. There are various types of complex networks because of different types of nodes and different types of edges. Taking a social network of people as an example, the nodes may represent people of different genders, nationalities, races, locations, ages, occupations, or other characteristics. The edges may represent various relationships such as friendship, colleague, schoolmate, romance, neighbour, or family. They can carry weights representing how relationships between two people close to each other, can be directed such that child $i$ may choose child $j$ as a friend, but child $j$ may not choose child $i$ as a friend, and can be signed representing trusted and untrusted relationships. Furthermore, there may be more than one type of edges in a network, such as friendship and family in a social network of people. Complex networks composed of weighted, directed, signed and multiple types of edges are called weighted[12], directed[13], signed[14,15] and multilayer networks[16,17], respectively. In this paper, we only focus on complex networks with a single type of node and a single type of edge. For convenience, we use complex networks to denote them unless otherwise specified.

To understand the formation, evolution and function of complex networks, it is crucial to explore their internal structures (also called structural regularities), including assortative structure, disassortative structure[18] and mixture structure[19], which have been widely applied in various practical systems such as recommendation systems based on social networks[20,21] and evolutionary games based on interdependent networks[22,23,24,25,26]. The assortative (i.e., community) structure, where most edges are within groups, and the disassortative (e.g., bipartite) structure, where most edges are across groups, are the two most common types of network structures. Beside them, the mixture structure is another type of network structure that is neither assortative structure nor disassortative structure, such as a mixture of them. The task of network structure exploration is to determine how many groups in a complex network (group number) and how to partition the nodes of the network into different groups (group partition).

During the past several years, a large number of structure exploration methods have been proposed for complex networks. Most of them are specialised for complex networks with only community structure and consider the exploration of community from different perspectives such as modularity[27] and random walks[28]. Modularity was first proposed by Girvan and Newman to assess the quality of a group partition, and was used for community structure detection[27]. Subsequently Clauset et al.[29] presented a hierarchical agglomeration model based on modularity, which used a fast greedy algorithm to optimize modularity. Schuetz and Caflisch[30] further extended the greedy algorithm using a multistep algorithm and a vertex mover refinement procedure for improvement. These three methods suffer from a resolution limit[31] which affects the quality of detected community. To solve this problem, Traag and Dooren[32] presented a resolution-free constant Potts model. Random walks that describe the observed behaviours of many processes have also been used for community structure detection. Pons and Latapy[33] introduced a similarity measure based on random walks to determine which nodes should be grouped together. Rosvall and Bergstrom[34] proposed an information-theoretic approach that used the probability flow of random walks as a proxy for information flows to reveal community structure. Recently, Bayesian nonparametric models[35] have also been used for community detection, which can automatically determine the group number. Morup and Schmidt[36] extended a Bayesian nonparametric model[37], originally used for clustering entities and relations of relational datasets, to automatically determine the group number in networks. Sinkkonen et al.[38] proposed a Bayesian nonparametric model to automatically determine the group number in large networks. A dilemma for all these methods is that the structure type of a network has to be prespecified, however, it is usually unavailable.

In order to analyse complex networks with any type of structure, Newman and Leicht[18] gave a general definition of structure where nodes connected with other nodes in similar patterns form groups and proposed a probabilistic mixture model to explore the structure. Wang and Lai[39] investigated the advantages and shortcomings of Newman's mixture model (NMM) and gave some suggestions for improvement. Shen et al.[40] presented a stochastic block-based model to show clear information on structural regularities in complex networks. The main disadvantage of these structural regularity exploration models is that specifying the group number is necessary.

In the real world, however, not only the group number but also the type of structure in a network is usually unknown in advance. To explore structural regularities in complex networks automatically, we extends the NMM model using Bayesian nonparametric theory and propose a novel Bayesian nonparametric model, called the Bayesian nonparametric mixture (BNPM) model. On the one hand, this model can explore general structural regularities in networks like the NMM model. On the other hand, it can automatically determine the group number because of Bayesian nonparametric theory. Experiments conducted on a large number of complex networks with different structures show that the BNPM model is able to automatically explore structural regularities in networks with a stable and state-of-the-art performance.

**Methods**

Generally, a network with $N$ nodes and $M$ edges is represented by an adjacency matrix $A$ of dimension $N \times N$, where $A_{ij} = 1, (1 \leq i, j \leq N)$ if there is a link from node $i$ to node $j$ and 0 otherwise. We use $N(i)$ to denote the out links of node $i$, that is, a set of neighbour edges of node $i$ in an undirected network. The task of network structure exploration is to determine the group number and group partition. In this section, we first introduce the NMM model, and then the BNPM model derived from the NMM model using Bayesian nonparametric theory. For convenience, we only introduce them on directed networks in detail.

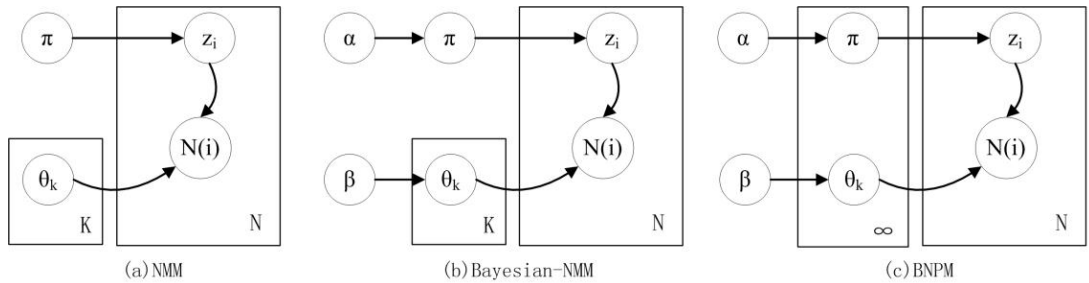

Figure 1 | Architectures of different mixture models.

*Newman's mixture model (NMM)*

A directed network with $K$ groups ($K$ is a predefined number) can be generated by the NMM model with two parameters $\pi_k$ and $\theta_{kj}$, where $\pi_k$ denotes the probability of a node in group $k$ ($k \in \{1,2,...,K\}$) and $\theta_{kj}$ denotes the probability of a link from a node in group $k$ connecting to node $j$, subjected to the normalization constraint $\sum_{j=1}^{N}\theta_{kj}=1$. The vector $\theta_k$ represents the characteristic of nodes in group $k$ linking to other nodes. According to $\theta_k$, nodes connecting to other nodes in similar patterns are grouped together. A network is generated in the following way (see Fig. 1(a)): for each node $i$ and its links $N(i)$, 1) Node $i$ falls into a group $z_i$ with probability $\pi_{z_i}$; 2) Each link $A_{ij} \in N(i)$ selects node $j$ with probability $\theta_{z_i j}$. The probability of a network $A$ with $N$ nodes can be written as

$$p(A,z\mid\pi,\theta)=p(A,z\mid\theta)\cdot p(z\mid\pi)=\prod_{i=1}^{N}\left(\pi_{z_i}\cdot\prod_{j=1}^{N(i)}\theta_{z_i j}^{A_{ij}}\right) \quad (1)$$

where $z_i \in \{1,...,K\}$ is a hidden variable that needs to be inferred. The logarithm of the equation (1):

$$\ln p(A,z\mid\pi,\theta)=\sum_{i=1}^{N}\left(\ln\pi_{z_i}+\sum_{j=1}^{N(i)}A_{ij}\ln\theta_{z_i j}\right) \quad (2)$$

Because of hidden variables $z_i, i=1,...,N$, parameters $\pi_k$ and $\theta_{kj}$ cannot be estimated using likelihood maximization estimation. Newman and Leicht used an expectation-maximization (EM)[41] algorithm to estimate them. In the E step, the expectation of equation (2) is:

$$\bar{L}=\sum_{i,k}q_{ik}\left[\ln\pi_k+\ln\sum_{j}A_{ij}\ln\theta_{kj}\right] \quad (3)$$

with

$$q_{ik}=\frac{\pi_k\prod_j\theta_{kj}^{A_{ij}}}{\sum_k\pi_k\prod_j\theta_{kj}^{A_{ij}}}, \quad (4)$$

where $q_{ik}$ denotes the probability of node $i$ belonging to group $k$. In the M step, the parameters $\pi$ and $\theta$ can be re-estimated by optimizing equation (3) as

$$\pi_k=\frac{\sum_i q_{ik}}{N}, \quad (5)$$

$$\theta_{kj} = \frac{\sum_i A_{ij} q_{ik}}{\sum_j \sum_i A_{ij} q_{ik}}, \tag{6}$$

*Bayesian nonparametric mixture (BNPM) model*

In order to introduce Bayesian nonparametric theory into NMM, we first extend the NMM model to Bayesian NMM by adding prior probabilities to $\pi$ and $\theta$, denoted as $\alpha$ and $\beta$ respectively as shown in Fig. 1(b). Using the Dirichlet distribution, the prior probabilities can be written as:

$$\pi \sim Dirichlet(\alpha), \tag{7}$$

$$\theta_k \sim Dirichlet(\beta), \tag{8}$$

Then the probability of a network $A$ with $N$ nodes is (refer to equation (1)):

$$p(A, z \mid \alpha, \beta) = p(A, z \mid \pi, \theta) p(\pi \mid \alpha) P(\theta \mid \beta) \tag{9}$$

Due to the conjugacy between the Dirichlet and Multinomial distributions, equation (9) can be simplified as:

$$p(A, z \mid \alpha, \beta) = p(A \mid z, \beta) p(z \mid \alpha) \tag{10}$$

with

$$p(z \mid \alpha) = \int p(z \mid \pi) p(\pi \mid \alpha) d\pi \tag{11}$$

$$p(A \mid z, \beta) = \int p(A \mid z, \theta) p(\theta \mid \beta) d\theta \tag{12}$$

Furthermore, we propose a nonparametric method to estimate the number of latent groups, and extend the Bayesian NMM model to BNPM (see Fig. 1(c)). The nonparametric method assumes that the number of latent groups is infinite, but only finite groups among them are used to generate any observed network. In the BNPM model, the probability of a network $A$ with $N$ nodes is (refer to equation (10)):

$$p(A, z, K \mid \alpha, \beta) = p(A \mid z, \beta) p(z, K \mid \alpha), \tag{13}$$

where $p(z, K \mid \alpha)$ is described as a Chinese Restaurant Process (CRP)[42] below:

$$p_{CRP}(z_i = k \mid z_1, ..., z_{i-1}) = \begin{cases} \dfrac{\alpha}{i-1+\alpha} & k \text{ is a new group} \\ \dfrac{n_k}{i-1+\alpha} & n_k > 0 \end{cases}, \tag{14}$$

where $n_k$ denotes the number of nodes already assigned to group $k$, and $\alpha$ is a hyperparameter. Equation (14) is the probability of assigning a new node to a new group or an existing group, and is free of the order of nodes.

In summary, a network is generated by the BNPM in the following way: 1) For each node $i$, assign it to a group $z_i$ according to $CRP(\alpha)$; 2) For each group $k$, draw the probability vector of nodes in it connecting to other nodes $\theta_k$ from $Dirichlet(\beta)$; 3) For each link of nodes $A_{ij} \in N(i)$, generate it according to the Multinomial distribution $Multinomial(\theta_{z_i,j})$ subjected to the normalization constraint $\sum_{j=1}^{N} \theta_{z_i,j} = 1$.

The hyperparameter $\alpha$, the priori probability of $\pi$, impacts on the number of groups. Although the group number is potentially infinite, the CRP gives an extremely uneven distribution over groups and ensures that the number of groups $K$ is much smaller than the number of nodes $N$ with an appropriate small value $\alpha$. The hyperparameter $\beta$, the prior probabilities of $\theta$, describes the degree distribution of a node within groups. To make the BNPM model flexible, we assume both $\alpha$ and $\beta$ follow Gamma distributions:

$$\alpha \sim G(1,1), \qquad \beta \sim G(1,1) \tag{15}$$

**Inference.** It is intractable to exactly estimate the latent variable $z$, so we employ Markov Chain Monte Carlo (MCMC) methods[43] for an approximate estimation. In the MCMC methods, Gibbs sampling[43] and slice sampling[44] are used to obtain samples of $z$ and hyperparameters ($\alpha$ and $\beta$), respectively.

**Sampling $z$** For each node $i$, given the group assignments of all other nodes $z_{-i}$, the probability of node $i$ choosing group $k$ is as follows (detailed information is presented in the supplementary data):

$$p(z_i = k \mid z_{-i}, A) = \prod_{j=1}^{N(i)} \frac{m_{k,-i}^j + \beta}{m_{k,-i} + N\beta + j - 1} \cdot \frac{F(n_k, \alpha)}{N + \alpha}, \tag{16}$$

where $F(n_k, \alpha) = n_k$, if $n_k > 0$; otherwise $F(0, \alpha) = \alpha$, meaning that a new group is generated. Here, $m_{k,-i}$ denotes the number of out links from nodes in group $k$ except node $i$, and $m_{k,-i}^j$ denotes the number of out links from nodes in group $k$ except node $i$ to node $j$. The initial group of each node is randomly assigned, and the data is sampled after the likelihood (equation (13)) reaches a stationary state.

**Hyperparameters** The optimal $\alpha$ and $\beta$ are selected from (0,1).

**Extension to undirected networks.** The BNPM model can also be extended to handle undirected networks like the NMM model. In an undirected network, when node $i$ in $z_i$ connects to another node $j$, node $j$ in $z_j$ also connects to node $i$. Again, we use $\theta_{z_i,j}$ to denote the probability of a node in group $z_i$ connecting to node $j$. $A_{ij} \in N(i)$ is generated according to the Multinomial distribution $Multinomial(\theta_{z_i,j}\theta_{z_j,i})$, where $\theta_{z_i,j}\theta_{z_j,i}$ denotes the probability of an edge between node $i$ and $j$, given nodes $i$ and $j$ in groups $z_i$ and $z_j$, respectively. It is subjected to the normalization constraint $\sum_{i=1}^{N}\sum_{j=1}^{N}\theta_{z_i,j}\theta_{z_j,i} = \left[\sum_{i=1}^{N}\theta_{z_j,i}\right]\cdot\left[\sum_{j=1}^{N}\theta_{z_i,j}\right] = 1$. As both $z_i$ and $z_j$ vary from 1 to $K$, $\sum_{i=1}^{N}\theta_{z_j,i} = \sum_{j=1}^{N}\theta_{z_i,j}$. Thus, $\sum_{j=1}^{N}\theta_{z_i,j} = 1$, exactly the same for directed networks. The remainder of the generative process follows the same way as before.

## Results

In this study, we first test the BNPM model on eight small networks with different structure types in detail, and then compare it with other related models on seven large networks besides the previous eight ones. Three out of the fifteen networks are synthetic networks and the others are real-world networks from different disciplines. Within these networks, the number of nodes ranges from tens to tens of thousands, and the number of edges ranges from tens to millions. Their detailed information is shown in Table 1. Among the seven large networks, we introduce Steam and Syn-10000 in detail as they have not been used before. The Steam network collected from store.steampowered.com has community structure to describe the friendship between game players. It consists of 39,252 players with 1,258,237 friendship edges. All players are partitioned into 7 groups according to the game they played during the recent two weeks: dota2, team fortress2, doomII, terraia, red faction armageddon, counter-strike: global offensive and NBA 2K15. The Syn-10000 network contains 10,000 nodes connected by 300,000 edges, where all the nodes are equally divided into 100 groups, and the edges are placed uniformly at random within and between them in certain numbers. For the first 40 groups, the number of edges in each group is set to 2,400, and the number of edges between a group and the others is set to 600. They form a community structure. The remaining 60 groups are further divided into 30 pairs, the number of edges between the two groups in each pair is set to 4,800, and the number of edges between groups in different pairs is set to 1,200. Each pair of groups forms a bipartite structure.

**Complex networks with only community structure.** Karate[45] and Dolphin[46], two undirected real-world complex networks that contain only community structure, are used to test the capability of the BNPM model to explore structural regularities in complex networks with only community structure. The Karate network characterises the acquaintance relationships among 34 members in the Zachary club, where all members (i.e., nodes) are divided into two groups, forming a community structure, due to a dispute between the administrator and karate teacher. The Dolphin network describes the frequent associations among 62 dolphins living off Doubtful Sound, New Zealand, where a dolphin connects to another dolphin when and only when they are observed together more often than expected by chance. The groups generated by the BNPM model on these two networks are shown in Fig. 2(a) and 2(b), respectively, where the nodes of the same shape are in the same gold group and the nodes of the same colour are in the same group generated by the model. The structural regularities in both networks automatically explored by the BNPM model are correct. For example, on the Karate network, not only the group number is correctly assigned as two, but also all the nodes are correctly grouped. For the sake of consistency, we also use shape and colour to distinguish the gold groups and the groups generated by the BNPM model in a complex network respectively in the following sections unless otherwise specified.

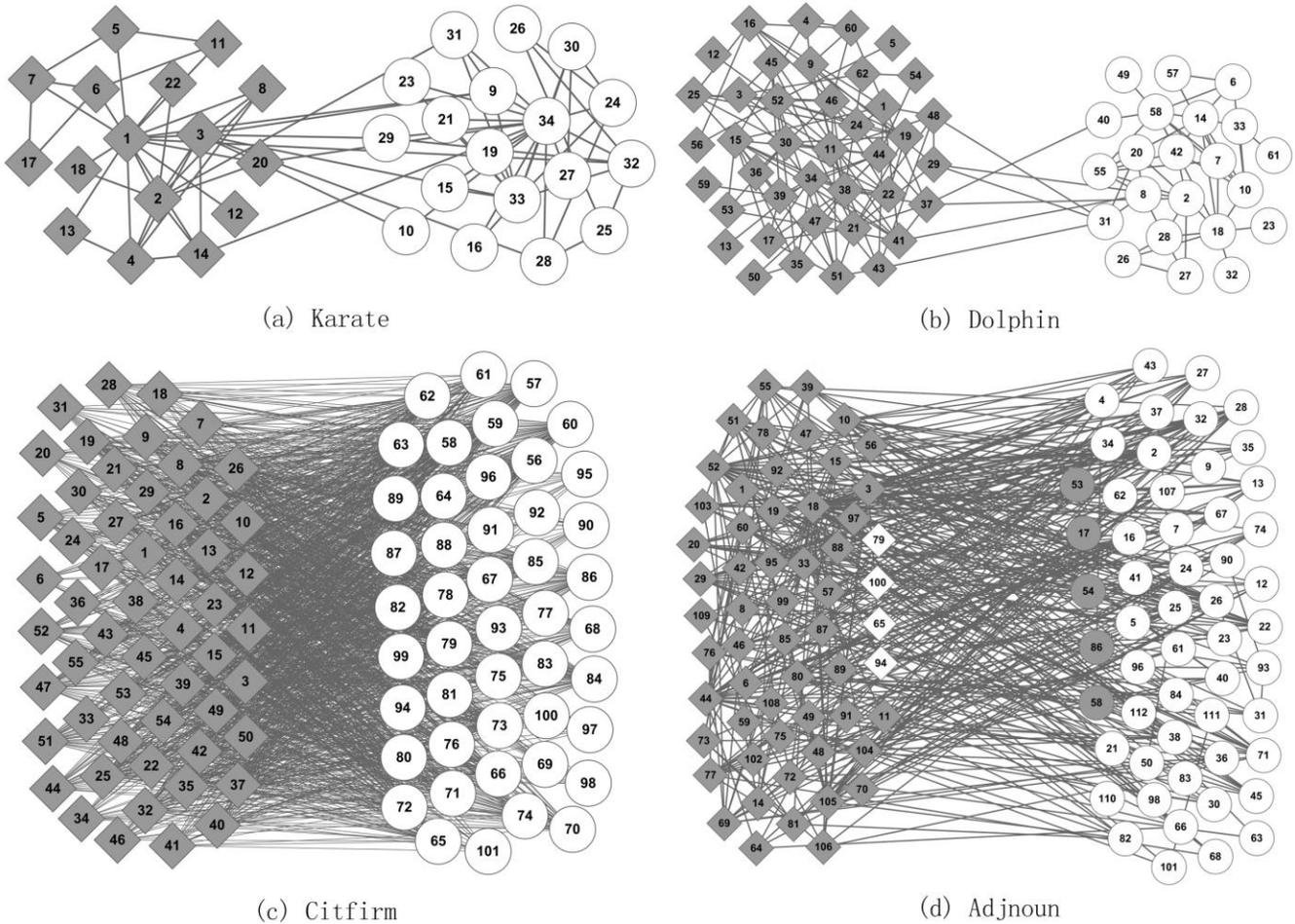

Figure 2 | Structural regularities explored by the BNPM model on the four complex networks that contain only community structure or bipartite structure: (a) Karate, b) Dolphin, c) Citfirm and d) Adjnoun. The nodes of the same shape are in the same gold group and the nodes of the same colour are in the same group generated by the BNPM model.

**Complex networks with only bipartite structure.** Citfirm[47] and Adjnoun[48], two undirected real-world networks that contain only bipartite structure, are used to test the capability of the BNPM model to explore structural regularities in complex networks with only bipartite structure. The Citfirm network describes a distribution of offices of 46 global advanced producer service firms over 55 cities in the world. The nodes of this network are divided into two groups: firm and city. The groups generated by the BNPM model on this network are shown in Fig. 2(c), where all the nodes (cities marked by diamonds and firms marked by circles) are automatically partitioned into two correct groups. The Adjnoun network consists of 112 common adjectives and nouns in the novel *David Copperfield* written by Charles Dickens, where the adjacent words connect with each other. As adjectives are always near the noun they are describing in English, the adjectives and nouns in the Adjnoun network form a bipartite structure. Figure 2(d) shows the groups generated by the BNPM model. All the nodes (adjectives marked by diamonds and nouns marked by circles) are automatically partitioned into two groups; one (black) is composed of most adjectives and the other one (white) is composed of most nouns. Four out of 58 adjectives {65, 79, 94, 100} and five out of 54 nouns {17, 53, 54, 58, 86} are wrongly grouped.

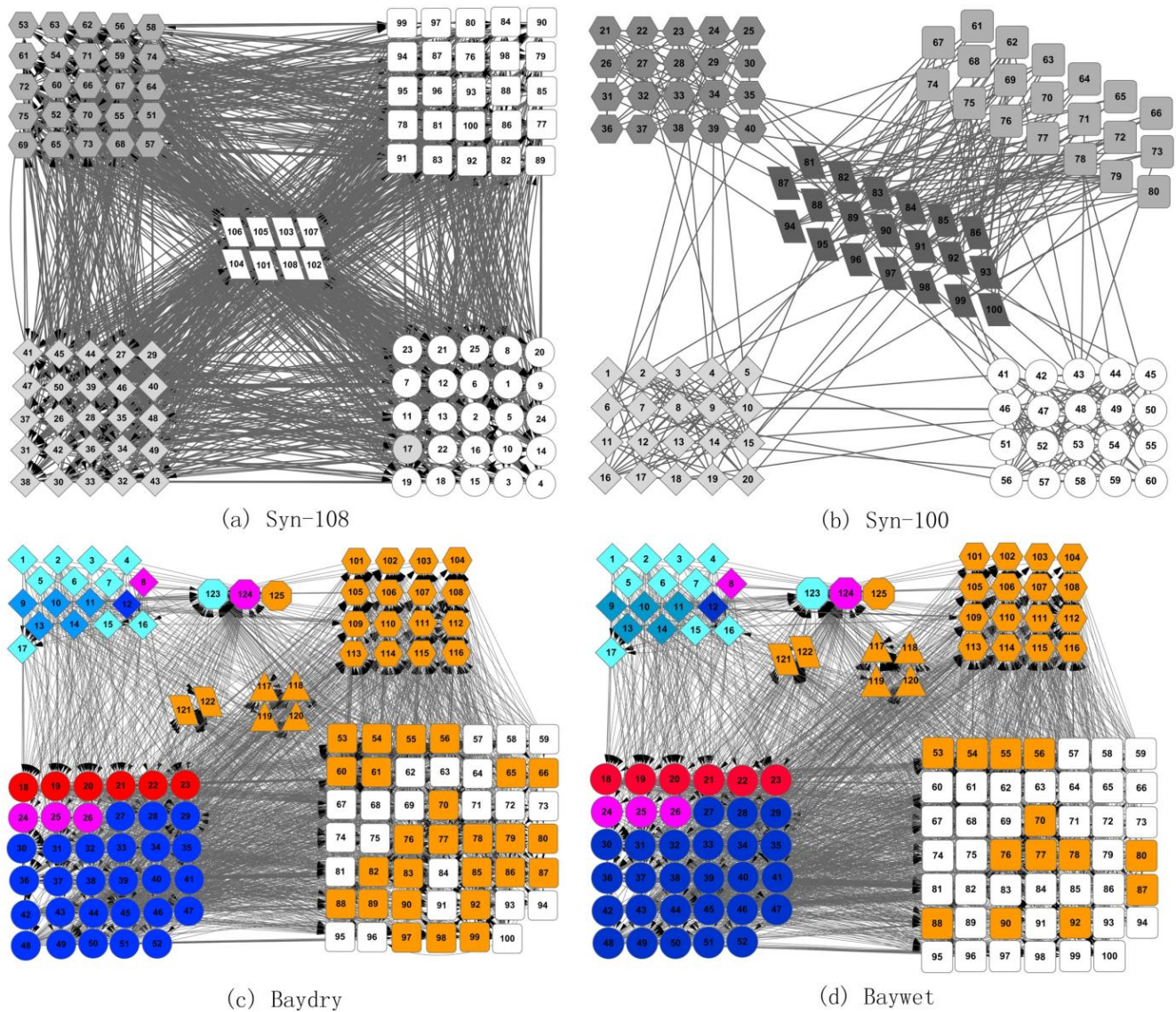

Figure 3 | Structural regularities explored by the BNPM model on the four complex networks with mixture structure: (a) Syn-108, b) Syn-100, c) Baydry and d) Baywet.

**Complex networks with mixture structure.** Syn-108[18] is a synthetic directed network composed of 108 nodes with 1439 edges provided by Newman. It contains a small number of keystone nodes that do not belong to any group but are able to guide other nodes to form different groups. In this network, 100 nodes are equally split into four groups, denoted A, B, C, and D, and uniformly link with each other at random using directed edges with a mean degree of 10. The nodes in these four groups link to 4 out of the remaining 8 nodes, called keystone nodes, according to their group membership in a circular manner. For instance, the nodes in groups A, B, C, and D link to the sets of keystone nodes {101,102,103,104}, {103,104,105,106}, {105,106,107,108}, and {107,108,101,102}, respectively, as shown in Fig. 3(a). In this way, no keystone node is uniquely identified with any group, but each group has a unique signature set of keystones, which is the only pattern to distinguish the group. Figure 3(a) shows the groups generated by the BNPM model on the Syn-108 network. All the nodes are automatically partitioned into three groups; one (the group in the upper left) is completely correct, another one (the group in the lower left) is almost completely correct except that one node (i.e., node 17) is wrongly grouped, and the remaining group contains all other nodes. Overall, 75 out of 100 nodes are partitioned into the correct groups, despite one absent group.

Syn-100 is a synthetic undirected network that has both community structure and bipartite structure. The network consists of 100 nodes with 402 edges as shown in Fig. 3(b), where the nodes fall into five groups, three of which form a community structure, whereas the other two groups form a bipartite structure. Figure 3(b) shows the groups generated by the BNPM model. All the nodes are automatically partitioned into five right groups.

Baydry and Baywet[49] are two directed networks that describe the real-world food webs of an ecosystem of 125 components in Florida Bay during the wet and dry seasons, where the nodes of both networks are divided into 7 compartments as shown in Fig. 3(c) and 3(d): Primary Producers (diamond), Invertebrates (ellipse), Fishes (round rectangle), Birds (hexagon), Reptiles (triangle), Mammals (parallelogram) and Detrital Compartments (octagon). On these two networks, the groups generated by the BNPM model are shown in Fig. 3(c) and 3(d), respectively. All the nodes in both networks are automatically partitioned into 7 groups, and the group partitions of the components from all compartments are the same except the Fishes. In particular, most Ivertebrate components are grouped in one group, and all Bird components are grouped into another group. As a whole 74 and 87 out of 125 nodes are correctly grouped on the two networks, respectively.

**Comparison with other related models.** We compare the BNPM model with other 9 related models on the 15 complex networks listed in the table 1. The related models fall into two categories: (C-I) models special for community structure detection, such as the Fastgreedy model[29], the Multi-Step Greedy (MSG) model[30], Traag's model[32], the Walktrap model[33], the Infomap model[34], the Bayesian community detection (BCD) model[36], and Sinkkonen's model[38]; (C-II) mixture models that require a specified group number for general structural regularity exploration, such as the Generalised Stochastic Block (GSB) model[40] and Newman's Mixture Model (NMM)[18]. In this study, we set the group number to the gold standard when applying the models in the second category (C-II). As the source codes of all the models except Sinkkonen's model have been released by their authors, we use the released source codes and implement Sinkkonen's model by ourselves.

The performance of each model is measured by the Normalised Mutual Information (NMI)[50],

$$P_{nmi}(G,G') = \frac{2MI(G,G')}{H(G)+H(G')}, \qquad (17)$$

where $G=(G_1,G_2,...,G_K)$ are the gold groups in a network, $G'=(G'_1,G'_2,...,G'_K)$ are the groups explored by an algorithm, $H(x)$ is the entropy of group $x$, and $MI(G,G')$ is the mutual information between $G$ and $G'$. The higher $P_{nmi}$, the better detection; specially, $P_{nmi} = 1$ means a perfect exploration.

Table 2 shows the $P_{nmi}$s of different models on the 15 networks, where NA denotes that a model is not applicable on a network or a model can not iterate one time within 24 hours. The BNPM model achieves the highest $P_{nmi}$s on most of the networks (13 out of 15), including 4 perfect $P_{nmi}$s on two networks with only a community structure (NID 1 and NID 2), a network with only a bipartite structure (NID 3) and a synthetic mixture network (NID 6). Compared with the models in C-I, the BNPM model achieves better $P_{nmi}$s on all the five networks (NID 1, NID 2, NID 9, NID 10 and NID 11) with only a community structure except NID 10. On NID 10, the $P_{nmi}$ of the BNPM model is comparable with the best $P_{nmi}$ of the models in C-I. Compared with the models in C-II, the BNPM model shows better performance on all the networks. Within both C-I and C-II, some models are not applicable on the large networks (NID 9-15). For example, the Fastgreedy model is not applicable on NID 12 and NID 15, and the NMM model is not applicable on NID 10-15.

In addition, we also check the group numbers determined by the BNPM model and compare them with that determined by the models in C-I. The results are shown in Table 3. The group numbers are correctly determined by the BNPM model on 7 out of 15 networks, including two networks with only a community structure (NID 1 and NID 2), two network with only a bipartite structure (NID 3 and NID 4) and three synthetic mixture network (NID 6, NID 7 and NID 8). Among them, only four networks are correctly partitioned as mentioned above. Compared with the models in C-I, the BNPM model shows some advantages; the BNPM model generates the best group numbers (nearest to the gold standards) on 8 out of the 15 networks, which is better than all the models in C-I.

**Discussions**

In this paper, we extend the NMM model using Bayesian nonparametric theory and propose a novel Bayesian nonparametric model, called the Bayesian nonparametric mixture (BNPM) model, to automatically explore structural regularities in complex networks. The BNPM model can determine not only the group number but also the group partition of different types of structures that are unknown in advance. Experiments conducted on a large number of complex networks with different structures show that the BNPM model is able to automatically explore structural regularities in networks with a stable and state-of-the-art performance.

Although the BNPM model achieves perfect results on four of the fifteen complex networks, it wrongly determines the group number or group partition on the other networks. We further analyse certain networks whose structural regularities are wrongly explored and find that the results of the BNPM are reasonable. For example, in the Adjnoun network, the nodes {65, 79, 94, 100} are assigned to the right group as shown in Fig. 2(d) because they connect only to the nodes in the left. This pattern corresponds exactly to the characteristic of bipartite structure that nodes within groups connect with each other more sparsely than they connect between groups. In the Baydry and Baywet networks shown in Fig. 3(c) and 3(d), the nodes {8, 24, 25, 26, 124} are grouped together because of the out links from them to the nodes in the Invertebrate compartment. It fits the fact that they are benthic components. Similarly, the BNPM model finds a group composed of zooplankton {18, 19, 20, 21, 22, 23}.

The reason why the NMM model is inferior to the BNPM model lies in that the NMM model suffers from group bias on some networks because of $q_{ik}=0$ in equation (4) whereas the BNPM model succeeds in avoiding this problem by adding the prior probabilities of $\pi$ and $\theta$. In the NMM model, given a network, when $q_{ik}=0$, node $i$ may not fall into group $k$, and will be wrongly assigned to another group. Specially, when $q_{i1}=...=q_{iK}=0$, node $i$ may not fall into any group, which will result in that the NMM model fails to handle this network. In the BNPM model, the $p(z_i=k|z_{-i},A)$ (see equation (16)) that has the same meaning of $q_{ik}$ is always greater than zero as $\frac{m_{k,-i}^j+\beta}{m_{k,-i}+N\beta+j-1} > \frac{m_{k,-i}^j}{m_{k,-i}+(N-1)\beta+j-1} \geq 0$. Just because of this, the NMM model is much inferior to the BNPM model on {NID 4, NID 5, NID 7, NID 8 and NID 9} and fails on NID10-15. Taking NID 4 and NID 14 as examples, node 5 and node 51 in NID 4 are assigned to wrong groups, and the nodes {1859, 5915, 7672, 8325, 9166, 9253} in NID 14 could not be assigned to any group.

The BNPM model shows state-of-the-art performance on various complex networks. However, it has certain limitations, for example, it is not currently applicable to weighted networks and multilayer networks. For further work, we plan to extend the BNPM model to be applicable to these networks.

**Acknowledgements**

This paper is supported in part by grants: NSFCs (National Natural Science Foundation of China) (61402128, 61473101, 61173075 and 61272383), Strategic Emerging Industry Development Special Funds of Shenzhen (JCYJ20140508161040764 and JCYJ201417172417105).

**Figure legends**

Figure 1 | Architectures of different mixture models.

Figure 2 | Structural regularities explored by the BNPM model on the four complex networks that contain only community structure or bipartite structure: (a) Karate, b) Dolphin, c) Citfirm and d) Adjnoun. The nodes of the same shape are in the same gold group and the nodes of the same colour are in the same group generated by the BNPM model.

Figure 3 | Structural regularities explored by the BNPM model on the four complex networks with mixture structure: (a) Syn-108, b) Syn-100, c) Baydry and d) Baywet.

**Tables**

Table 1 | Detailed information of the fifteen networks used in our study.

| NID | Name | $N$ | $M$ | $K$ | Directed | Structure type | Descriptions / Group nature |
|---|---|---|---|---|---|---|---|
| 1 | Karate | 34 | 78 | 2 | no | Community | Zachary's karate club[45] / Membership |
| 2 | Dolphin | 62 | 159 | 2 | no | Community | Dolphin social network[46] / Membership |
| 3 | Citfirm | 101 | 1,342 | 2 | no | Bipartite | World cities & global firms[47] / Location |
| 4 | Adjnoun | 112 | 425 | 2 | no | Bipartite | Word network[48] / Word characteristic |
| 5 | Syn-108 | 108 | 1,439 | 4 | yes | Mixture | Synthetic network[18] |
| 6 | Syn-100 | 100 | 402 | 5 | no | Mixture | Synthetic network |
| 7 | Baydry | 125 | 1,969 | 7 | yes | Mixture | Florida bay food-web[49] / Species |
| 8 | Baywet | 125 | 1,938 | 7 | yes | Mixture | Florida bay food-web[49] / Species |
| 9 | Yeast | 2,361 | 6,646 | 13 | no | Community | Protein interaction network[51] / Protein function |
| 10 | Openflight | 3,266 | 33,544 | 9 | no | Community | OpenFlight Airport database[52] / Tz database area |
| 11 | Steam | 39,252 | 1,258,237 | 7 | no | Community | Steam friendship network / Game type |
| 12 | Lederberg | 8,324 | 41,450 | 57 | no | Mixture | Citation network of Lederberg[53] / Published year |
| 13 | California | 8,722 | 14,080 | 7 | yes | Mixture | Webpage of California[54] / Domain suffix |
| 14 | Syn-10000 | 10,000 | 300,000 | 100 | no | Mixture | Synthetic network |
| 15 | Freeassoc | 10,617 | 72,168 | 10 | yes | Mixture | Word network[55] / Word characteristic |

Table 2 | $P_{nmi}$s of different models on the fifteen networks.

| NID | Models special for community structure detection (C-I) | | | | | | | Mixture models (C-II) | | Our model |
|---|---|---|---|---|---|---|---|---|---|---|
| | Fastgreedy | MSG | Traag's | Walktrap | Infomap | BCD | Sinkkonen's | GSB | NMM | BNPM |
| 1 | 0.6925 | 0.5515 | 0.5866 | 0.5618 | 0.6995 | 0.6882 | **1** | **1** | **1** | **1** |
| 2 | 0.6113 | 0.5345 | 0.4906 | 0.5254 | 0.5027 | 0.443 | **1** | 0.8904 | **1** | **1** |
| 3 | 0.0206 | 0.0043 | 0.0246 | 0.242 | 0 | 0.4009 | 0.0353 | **1** | **1** | **1** |
| 4 | 0.0025 | 0.0058 | 0.0036 | 0.0673 | 0.0315 | 0.0484 | 0 | 0.5032 | 0.5084 | **0.5967** |
| 5 | 0.1331 | 0.177 | 0.2006 | 0.3453 | 0 | 0.5837 | 0.2832 | 0.5118 | 0.683 | **0.8257** |
| 6 | 0.8786 | 0.8786 | 0.9057 | 0.8619 | 0.8753 | 0.8937 | 0.8313 | **1** | **1** | **1** |
| 7 | 0.0593 | 0.2343 | 0.2255 | 0.2471 | 0 | 0.516 | 0.2612 | 0.3744 | 0.4807 | **0.628** |
| 8 | 0.1134 | 0.3113 | 0.2572 | 0.331 | 0 | 0.519 | 0.2613 | 0.3564 | 0.4681 | **0.6716** |
| 9 | 0.152 | 0.1525 | 0.0786 | NA | 0.2611 | 0.0936 | 0.0441 | 0.0629 | 0.085 | **0.3674** |
| 10 | 0.5716 | 0.6154 | **0.6402** | 0.5518 | 0.5206 | 0.2744 | 0.505 | 0.3716 | NA | 0.5712 |
| 11 | 0.0983 | 0.1687 | 0.173 | NA | 0.2006 | 0.1973 | NA | 0.1412 | NA | **0.3193** |
| 12 | NA | 0.0971 | 0.1066 | 0.1412 | **0.2375** | 0.1336 | 0.0659 | 0.134 | NA | 0.1366 |
| 13 | 0.196 | 0.1931 | 0.167 | NA | 0.1896 | 0.027 | 0.0396 | 0.0241 | NA | **0.1994** |
| 14 | 0.7422 | 0.7514 | 0.8926 | 0.9469 | 0.9527 | 0.665 | 0.5483 | NA | NA | **0.9995** |
| 15 | NA | 0.0196 | 0.0237 | 0.1052 | 0.0555 | 0.2045 | 0.0213 | 0.018 | NA | **0.2978** |

Table 3 | Group numbers determined by the models in C-I and BNPM model on the fifteen networks.

| NID | Fastgreedy | MSG | Traag's | Walktrap | Infomap | BCD | Sinkkonen's | BNPM | Gold standard |
|---|---|---|---|---|---|---|---|---|---|
| 1 | 3 | 4 | 4 | 3 | 3 | 6 | **2** | **2** | 2 |
| 2 | 4 | 4 | 5 | 5 | 6 | 9 | **2** | **2** | 2 |
| 3 | 3 | 3 | 5 | 8 | 1 | 16 | 4 | **2** | 2 |
| 4 | 7 | 6 | 7 | 12 | 7 | 13 | 1 | **2** | 2 |
| 5 | 3 | 5 | 5 | **4** | 1 | 6 | 6 | 3 | 4 |
| 6 | 4 | 4 | 4 | 4 | **5** | 4 | **5** | **5** | 5 |
| 7 | 2 | 3 | 3 | 3 | 1 | 29 | 4 | **7** | 7 |
| 8 | 3 | 3 | 3 | 33 | 1 | 28 | 4 | **7** | 7 |
| 9 | 53 | 53 | 47 | NA | 202 | 29 | **9** | 26 | 13 |
| 10 | 49 | 39 | 26 | 114 | 135 | 40 | **8** | 19 | 9 |
| 11 | 4 | **5** | 17 | NA | 99 | 119 | NA | 69 | **7** |
| 12 | NA | 24 | 25 | 94 | 272 | **53** | 10 | 40 | 57 |
| 13 | 123 | 344 | 104 | NA | 423 | 28 | **5** | 17 | 7 |
| 14 | 18 | 19 | 47 | 70 | 70 | 18 | 10 | **101** | 100 |
| 15 | NA | 29 | 15 | 111 | 384 | 38 | **11** | 13 | 10 |